\documentclass{aip-cp}

\usepackage[numbers]{natbib}
\usepackage{rotating}
\usepackage{graphicx}
\usepackage{lineno}

\newcommand{\pta}{p_{\rm{T}}}
\newcommand{\pp}{pp}
\newcommand{\pPb}{p--Pb}
\newcommand{\Pbp}{Pb--p}
\newcommand{\PbPb}{Pb--Pb}

\begin{document}


\title{Ridges in p--A (and pp) collisions}

\author[aff1]{Alice Ohlson\corref{cor1}\\ for the ALICE, ATLAS, CMS, and LHCb Collaborations}

\affil[aff1]{CERN, Geneva, Switzerland}
\corresp[cor1]{Corresponding author: alice.ohlson@cern.ch}

\maketitle

\begin{abstract}
Correlations between particles separated by several units of pseudorapidity were discovered in high-multiplicity \pp{} and \pPb{} collisions at the LHC.  These long-range structures observed in two-particle correlation functions are reminiscent of features seen in \PbPb{} collisions, where they are often viewed as a signature of collective behavior and the formation of a quark-gluon plasma (QGP).  Therefore, the discovery of these `ridges' in small systems has implications for the study of collectivity in small systems as well as in heavy-ion collisions.  The ridges in \pp{} and \pPb{} collisions have been studied in the ALICE, ATLAS, CMS, and LHCb experiments to characterize the $\pta{}$-, $\eta$-, and multiplicity-dependences of the ridge yield, as well as its particle composition.  
\end{abstract}

\section{INTRODUCTION}
Two-particle angular correlations are used to study many aspects of the physics of heavy-ion collisions, in particular jet fragmentation and collective effects.  The correlation function is defined as the distribution in relative azimuthal angle ($\Delta\varphi = \varphi_{assoc}-\varphi_{trig}$) and relative pseudorapidity ($\Delta\eta = \eta_{assoc}-\eta_{trig}$) between trigger and associated particles.  In order to study various physical mechanisms, correlation functions can be constructed differentially as a function of, for example, the transverse momentum ($\pta{}$), species, and pseudorapidity of the trigger and associated particles, and the centrality or multiplicity of the collision.  

Correlation functions in \pp{} collisions show characteristic features attributed to jet production: a nearside peak localized around $(\Delta\varphi,\Delta\eta) = (0,0)$, representing pairs of particles where the trigger and associated particles are fragments of the same jet, and the awayside peak localized around $\Delta\varphi = \pi$ but extended in $\Delta\eta$, representing pairs in which the trigger and associated particles are in back-to-back jets.   In heavy-ion collisions, the same jet features are observed, on top of additional structures extended in $\Delta\eta$ around $\Delta\varphi = 0$ and $\Delta\varphi = \pi$.  These `ridges' are often viewed as a signature of collective behavior and the formation of a quark-gluon plasma, and are attributed to hydrodynamic flow in the QGP.  The bulk features in the two-particle correlation function (excluding the jet components) can be described by a Fourier series, 
\begin{equation}
\frac{dN}{d\Delta\varphi} \propto 1+2v_1^{trig}v_1^{assoc}\cos(\Delta\varphi)+2v_2^{trig}v_2^{assoc}\cos(2\Delta\varphi)+2v_3^{trig}v_3^{assoc}\cos(3\Delta\varphi)+...
\label{eq:vn}
\end{equation}
where the $v_2$ term is dominant in all but the most central collisions.  Measuring $v_2$ and the other $v_n$ components has been critical to determining the properties and dynamics of the medium created in heavy-ion collisions.  


Causality arguments suggest that the origin of correlations between particles separated in $\eta$ should be in the very early stages of the collisions, either in the initial state or in the initial energy distribution.  In the latter case, collective behavior is needed to transform the spatial correlations into the observed momentum-space correlations.  The discovery of long-range $\Delta\eta$ correlations (a ridge) on the nearside (around $\Delta\varphi = 0$) in high multiplicity collisions of small systems, \pp{}~\cite{ppridge} and \pPb{}~\cite{pPbridge}, therefore informs theories concerning the initial state of nuclear collisions and collective behavior in small and large systems.

\section{Ridges in \pPb{} collisions}

\begin{figure}[t!]
\centering \includegraphics[width=0.3\textwidth]{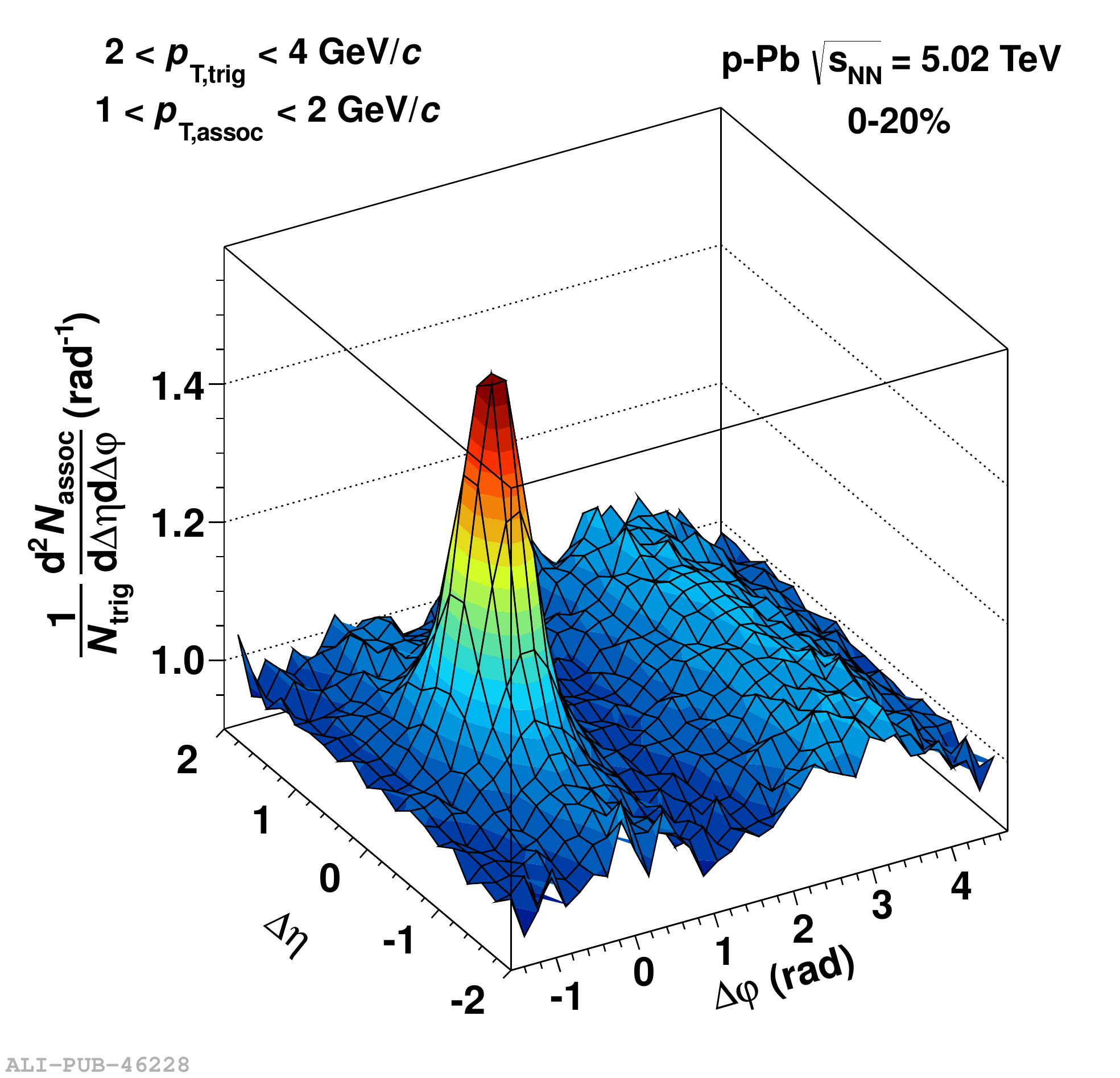}~
\centering \includegraphics[width=0.3\textwidth]{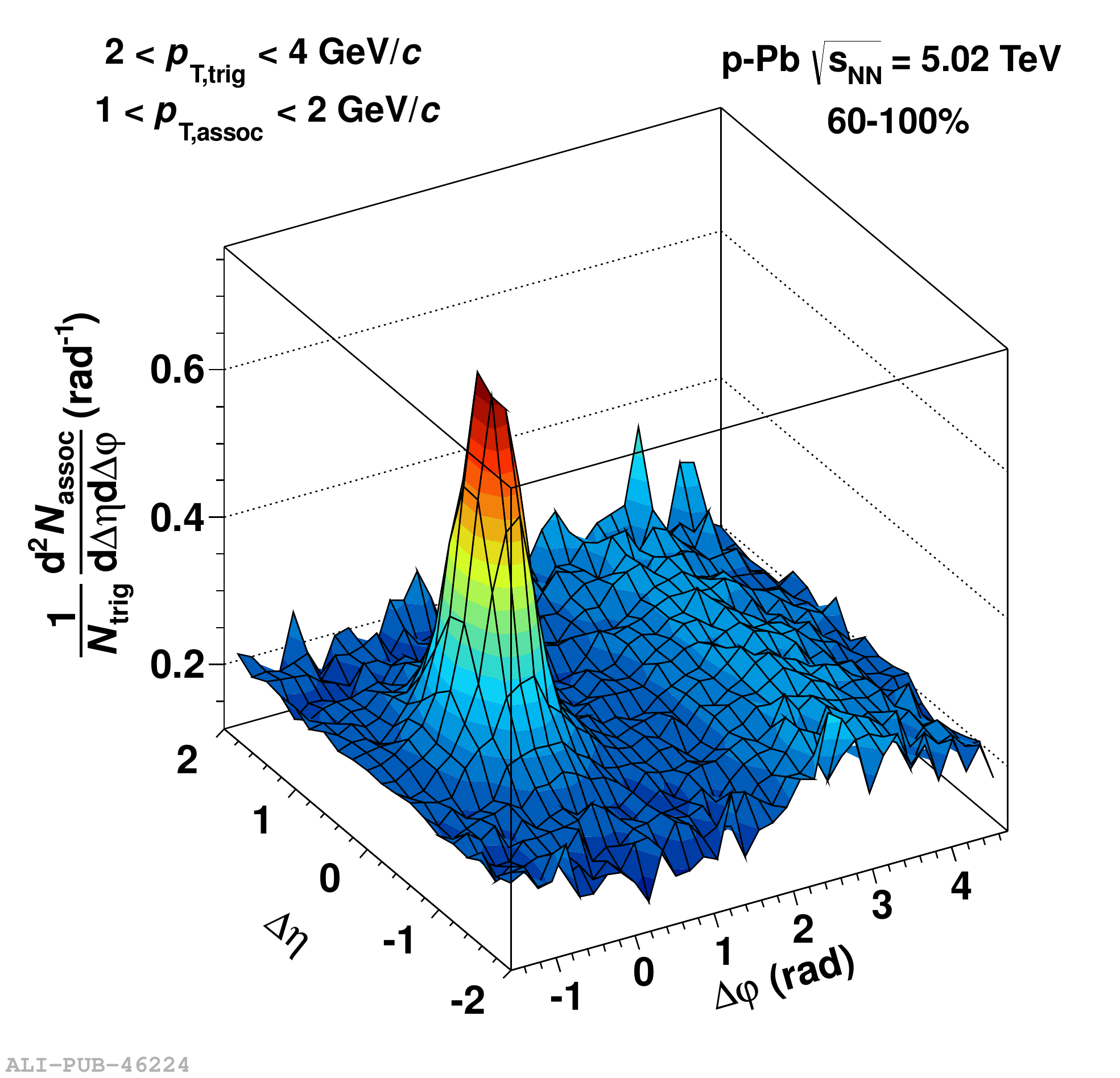}~
\centering \includegraphics[width=0.3\textwidth]{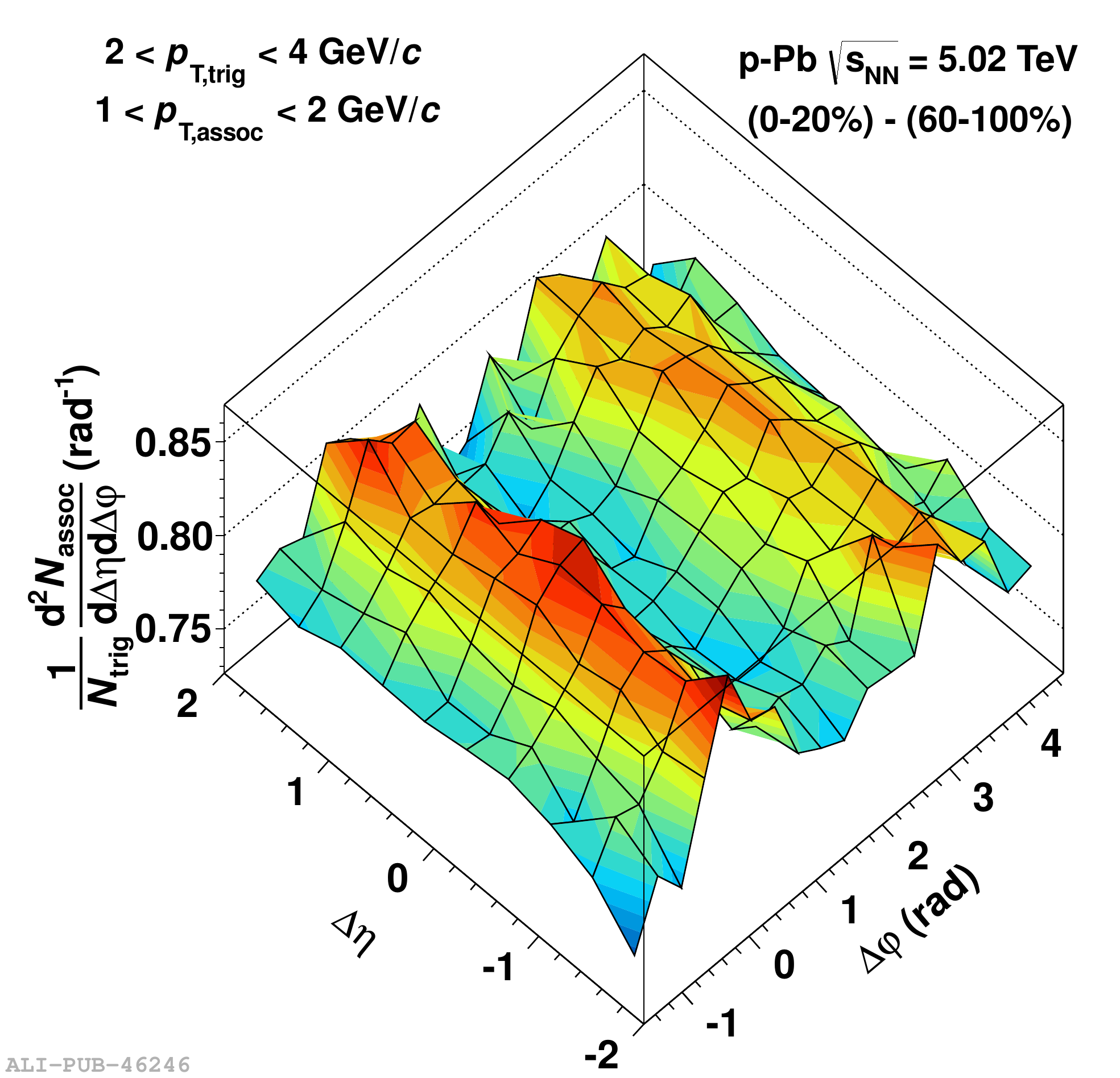}
\caption{(left) The two-particle correlation function in high-multiplicity \pPb{} collisions shows a ridge around $\Delta\varphi\sim 0$.  (center) The correlation function in low-multiplicity collisions shows no visible ridge.  (right) The subtracted distribution reveals the double ridge structure~\cite{doubleridge}.}\label{fig:doubleridge}
\end{figure}

The nearside ridge has been observed in high-multiplicity \pPb{} collisions by ALICE~\cite{doubleridge}, ATLAS~\cite{doubleridgeATLAS}, CMS~\cite{pPbridge}, and LHCb~\cite{lhcbRidge}.  Furthermore, it was observed that the nearside peak yields are mostly independent of multiplicity~\cite{minijet}, meaning that for the same trigger and associated $\pta{}$ the same jet population is selected regardless of multiplicity.  This served as justification to subtract the correlations in low-multiplicity events from the high-multiplicity correlation functions in order to remove correlations due to jet and minijet fragmentation.  This subtraction procedure, shown in Fig.~\ref{fig:doubleridge}, showed the nearside ridge more clearly and also revealed a symmetric ridge on the awayside~\cite{doubleridge,doubleridgeATLAS}, which is reminiscent of the ridges attributed to flow in heavy-ion collisions.  This `double ridge' structure was decomposed into Fourier coefficients (Eq.~\ref{eq:vn}) in order to extract the parameter $v_2$.  However, it is important to note that the physical mechanism leading to a non-zero $v_2$ is still under theoretical debate, and the presence of $v_2$ does \emph{not} necessarily imply the existence of hydrodynamics or a QGP in small systems.  

\begin{figure}[b!]
\centering \includegraphics[width=0.55\textwidth]{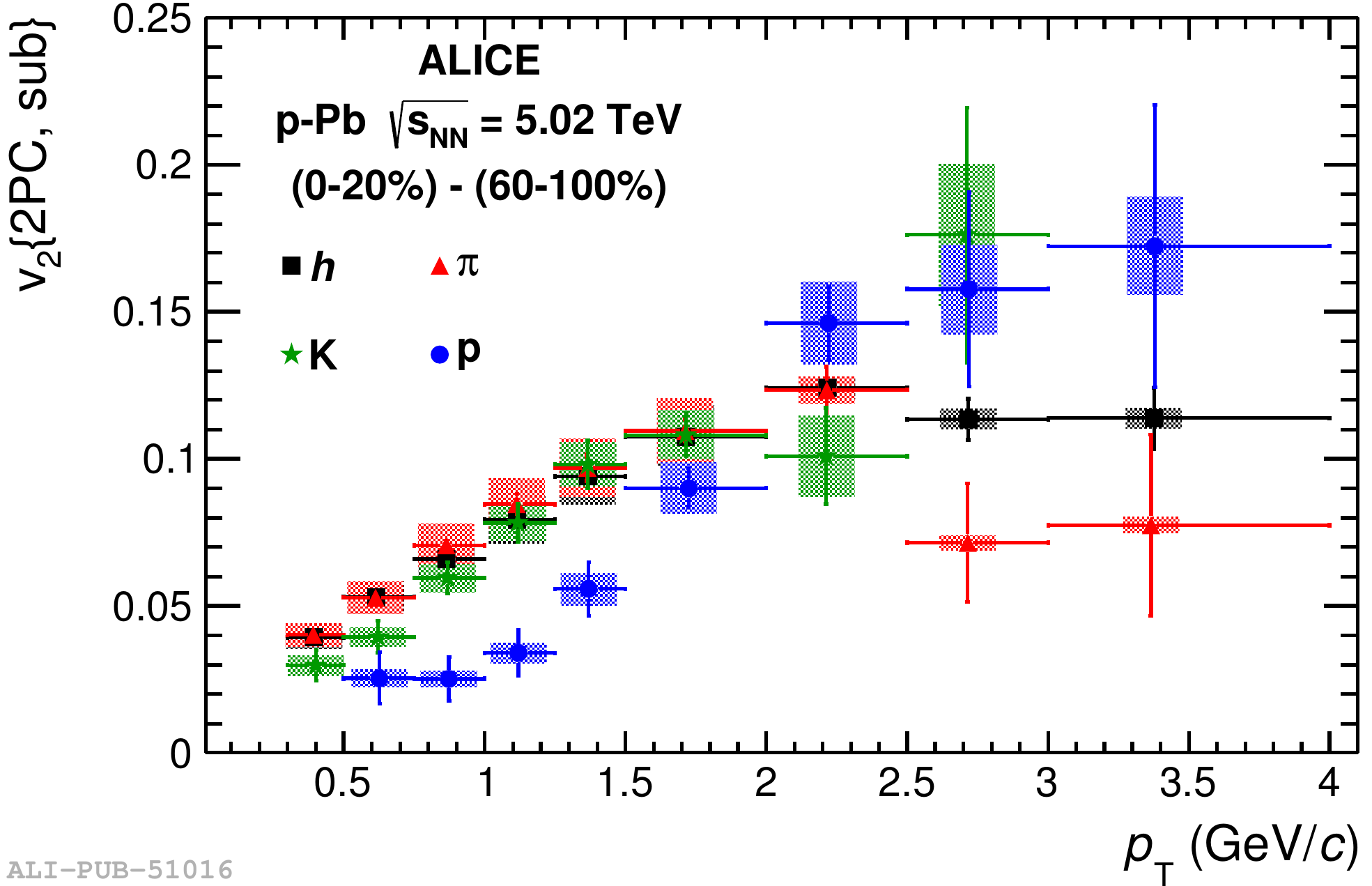}~
\centering \includegraphics[width=0.4\textwidth]{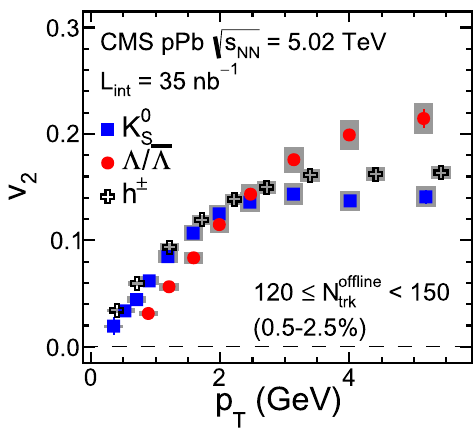}
\caption{(left) The $v_2$ of $\pi$, $K$, and $p$ as well as inclusive unidentified hadrons was measured as a function of $\pta{}$ in \pPb{} with the subtraction method~\cite{PIDridge}.  (right) The $v_2$ of $K^0_S$ and $\Lambda$ as well as inclusive unidentified hadrons was measured with the harmonic decomposition technique~\cite{cmsv2}.\label{fig:cmsv2}}
\end{figure}

Further studies in \pPb{} collisions measured $v_2$ as a function of $\pta{}$ for various particle species: $\pi^{\pm}$, $K^{\pm}$, $p(\bar{p})$, $K^0_S$, and $\Lambda(\bar{\Lambda})$.   The observed $v_2$, shown in Fig.~\ref{fig:cmsv2}, shows similar mass ordering as was observed in \PbPb{} collisions~\cite{PIDridge,cmsv2}.  Additionally, higher-order $v_n$ coefficients were measured, and shown to be non-zero up to $n=5$~\cite{ATLASvn}.  Finally, $v_2$ was measured using the multiparticle cumulant method, which measures multiparticle correlations while subtracting all lower-order correlations, as well as the Lee-Yang Zeros method~\cite{CMScumul}.  The agreement of the $v_2\{4\}$, $v_2\{6\}$, $v_2\{8\}$, and $v_2\mbox{\{LYZ\}}$ measurements demonstrates that the double ridge structure is a global correlation amongst many particles in a given event.  

Several theoretical explanations have been proposed to explain the ridge in \pp{} and \pPb{} collisions, such as higher-order glasma graphs within a Color Glass Condensate (CGC) picture, hydrodynamics in small systems, and others (see for example \cite{hydroridge,cgcridge}).  Since some models make different predictions for the $\eta$-dependence of the ridge, making such a measurement may enable us to differentiate between theories for the physical mechanism responsible for the ridge phenomenon.  To this end, several experiments have measured the $\eta$-dependence of the ridge.  CMS measures $v_2(\eta)$ at midrapidity up to $|\eta| = 2.4$ while ALICE and LHCb measure $v_2$ and the ridge at forward rapidity up to $\eta\sim 4$ and $\eta\sim 5$, respectively.  Because the LHC provided circulating proton and lead beams in both directions, the experiments were able to take data in both \pPb{} and \Pbp{} configurations, and measure observables in both the p-going and Pb-going directions with the same experimental setup and under the same conditions.  

\begin{figure}[t!]
\centering 
\includegraphics[width=0.55\textwidth]{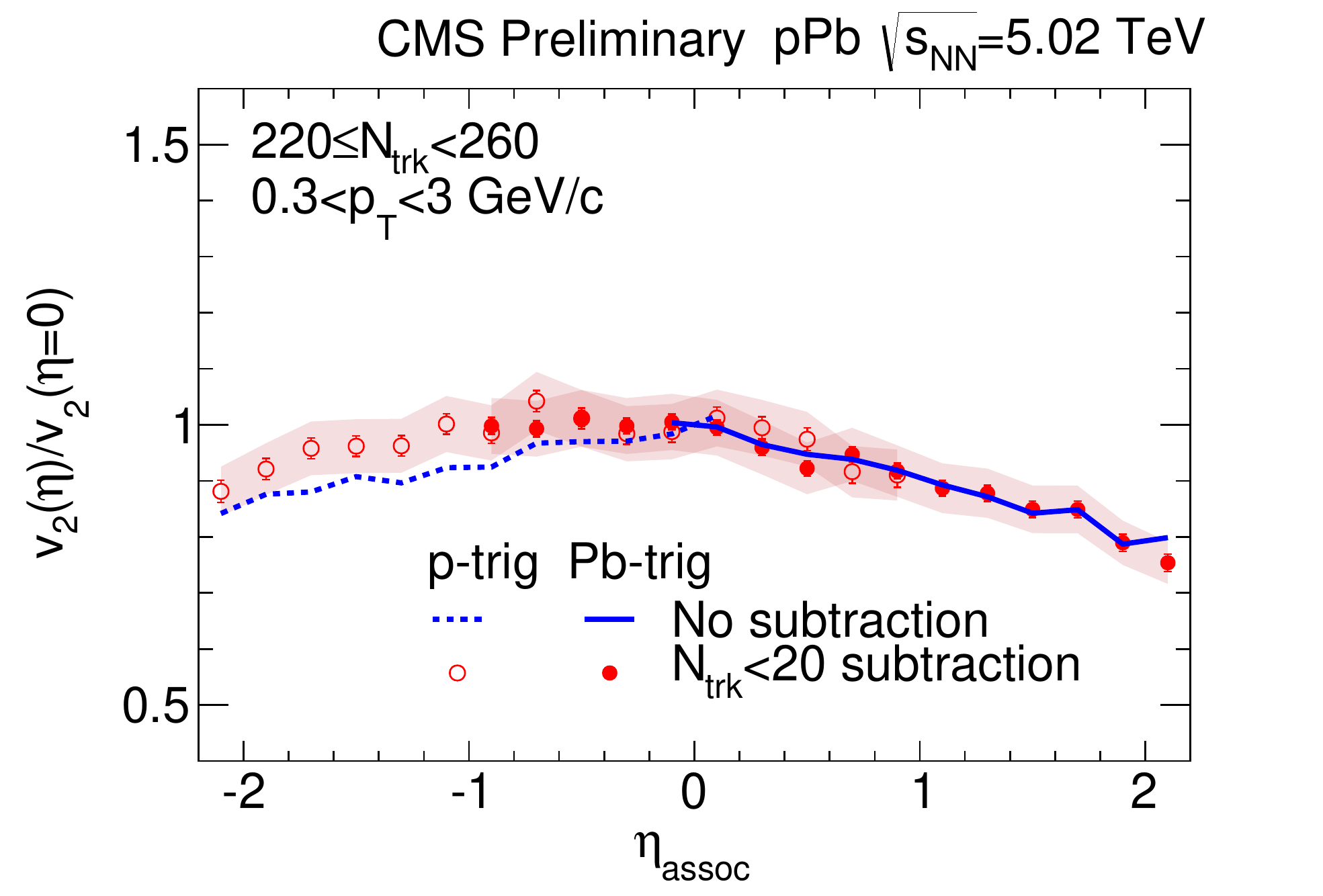}
\includegraphics[width=0.45\textwidth]{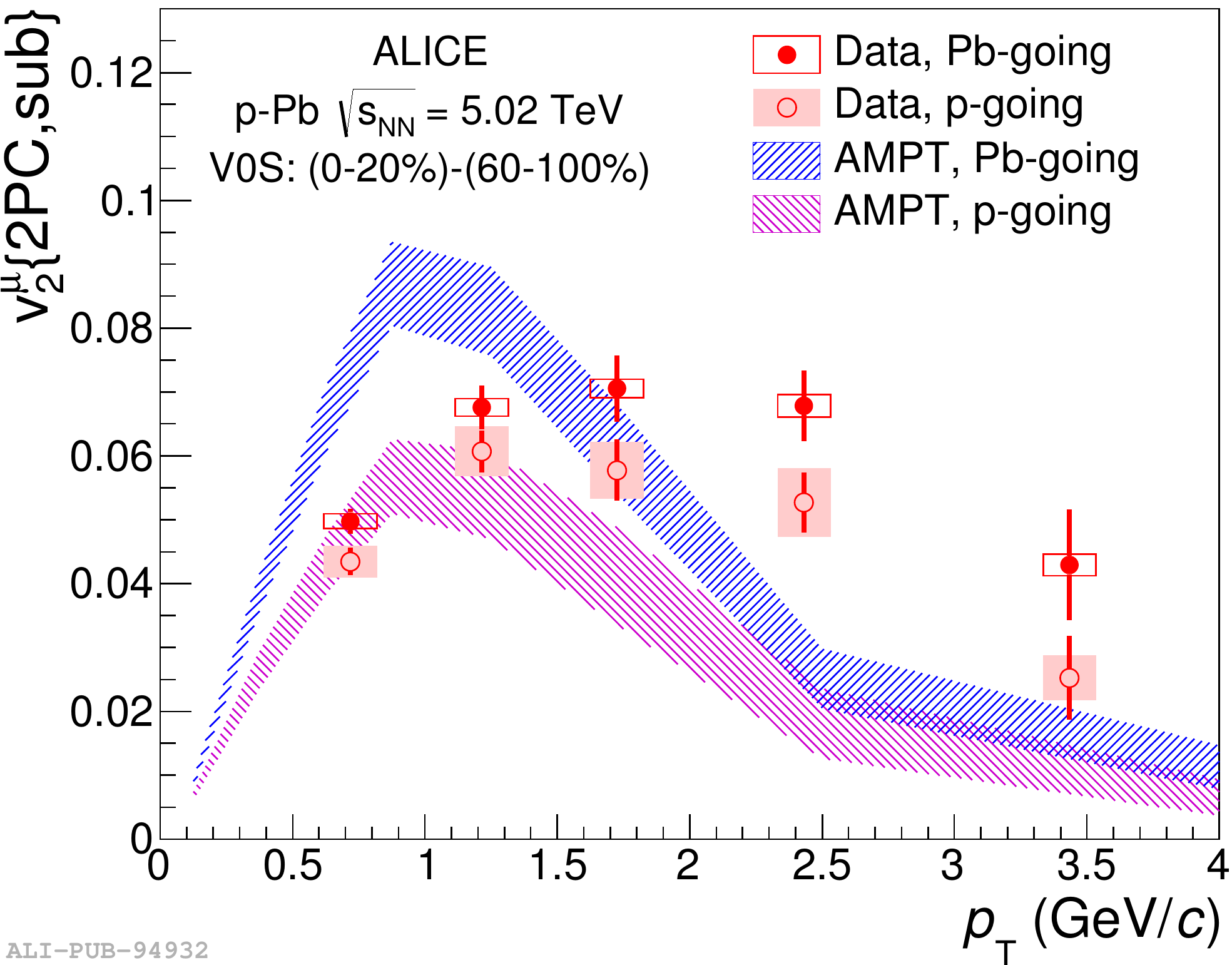}
\caption{(left) The $v_2$ of particles at midrapidity was measured as a function of pseudorapidity relative to $v_2(\eta=0)$~\cite{cmsEtaV2}.  (right) The $v_2$ of forward muons was measured as a function of $\pta{}$~\cite{aliceMuh}. Both results show that the $v_2$ is higher for particles traveling in the Pb-going direction than in the p-going direction.\label{fig:muh}}
\end{figure}

Fig.~\ref{fig:muh} shows a measurement from two-particle correlations where the trigger particle is fixed in the range $2.0 < |\eta_{trig}| < 2.4$ and the pseudorapidity of the associated particle ($\eta_{assoc}$) is varied.  In the ratio of $v_2^{trig}v_2^{assoc}$ measured at a given pseudorapidity to the $v_2^{trig}v_2^{assoc}$ at $\eta = 0$, the $v_2$ of the trigger particle cancels out, leaving only $v_2(\eta_{assoc})/v_2(0)$ as a function of $\eta_{assoc}$ as shown in the figure.  It is observed that the measured $v_2$ is larger for associated particles in the Pb-going direction than in the p-going direction.  

\begin{figure}[b!]
\centering \includegraphics[width=\textwidth]{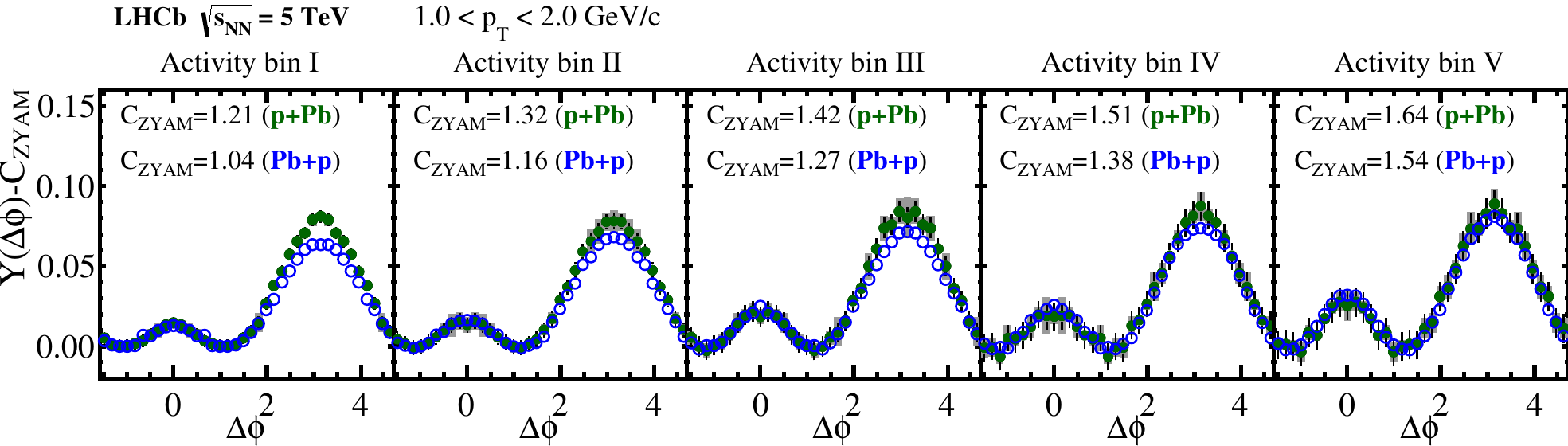}
\caption{Correlation functions between trigger and associated particles in the range $2.0 < |\eta| < 4.9$ show that the ridge yield is identical in both directions when compared in the same \emph{absolute} multiplicity bins~\cite{lhcbRidge}.\label{fig:lhcb}}
\end{figure}

The CMS finding was confirmed in the $\pta{}$-differential study of the $v_2$ of forward muons in ALICE.  In this analysis, the correlations between mid-rapidity charged hadrons reconstructed as `tracklets' in the Inner Tracking System (ITS, $|\eta| < 1$) and forward muons detected in the Forward Muon Spectrometer (FMS, $-4 < \eta < -2.5$) are constructed in the highest 20\% multiplicity events and the lowest 40\% multiplicity events.  The correlation functions in low-multiplicity events are subtracted from those in high-multiplicity events, and then fit with a Fourier series to extract $v_2^{\mu}\{\mbox{2PC,sub}\}$, which is shown as a function of $\pta{}$ in Fig.~\ref{fig:muh}.  Furthermore, the ratio of $v_2^{\mu}\{\mbox{2PC,sub}\}$ in the Pb-going direction to the $v_2^{\mu}\{\mbox{2PC,sub}\}$ in the p-going direction is roughly independent of $\pta{}$ within statistical and systematic uncertainties, and a constant fit to the ratio yields $1.16\pm0.06$.  The data were compared to an AMPT~\cite{refAMPT} simulation, which qualitatively describes the $\pta$-dependence at low $\pta$, but shows significant quantitative differences in both the $\pta$- and $\eta$-dependences.  

LHCb also confirms that for a given \emph{relative} event activity (i.e. mutiplicity percentile), the ridge is stronger in the Pb-going direction than in the p-going direction.  However, when comparing correlation functions in both beam configurations with the same \emph{absolute} event activity (defined by particle production in $2.0 < \eta < 4.9$), as shown in Fig.~\ref{fig:lhcb}, it is observed that the nearside ridge magnitude is the same in both hemispheres.  This new observation will serve to constrain models of $\eta$-dependence of the ridge yield.  

\section{Ridges in \pp{} collisions}
The nearside ridge in \pp{} was first observed by CMS in high multiplicity collisions at $\sqrt{s} = 7~\mbox{TeV}$, and further measurements showed that it was most prominent in the intermediate $\pta{}$ region ($1 \lesssim \pta{} \lesssim 3~\mbox{GeV}/c$)~\cite{ppridge}.  First results at a higher beam energy of $\sqrt{s} = 13~\mbox{TeV}$ once again clearly showed the existence of the nearside ridge~\cite{atlas13tev}.  Measurements performed in ATLAS of the ridge yield as a function of multiplicity and $\pta{}$ showed no significant differences between the two center-of-mass energies~\cite{atlas13tev}.  However, it should be noted that there are differences between the two analyses (for example, in the definition of the multiplicity $N_{ch}$ and the $|\Delta\eta|$ integration range) which prevent a precise comparison between the ATLAS and CMS results.  Despite significant challenges due to the influence of jets in \pp{} collisions, later work by ATLAS produced a measurement of $v_2$ in \pp{} collisions at $\sqrt{s} = 13~\mbox{TeV}$~\cite{atlas13tevNew}.  

\begin{figure}[t!]
\centering 
\includegraphics[width=0.47\textwidth]{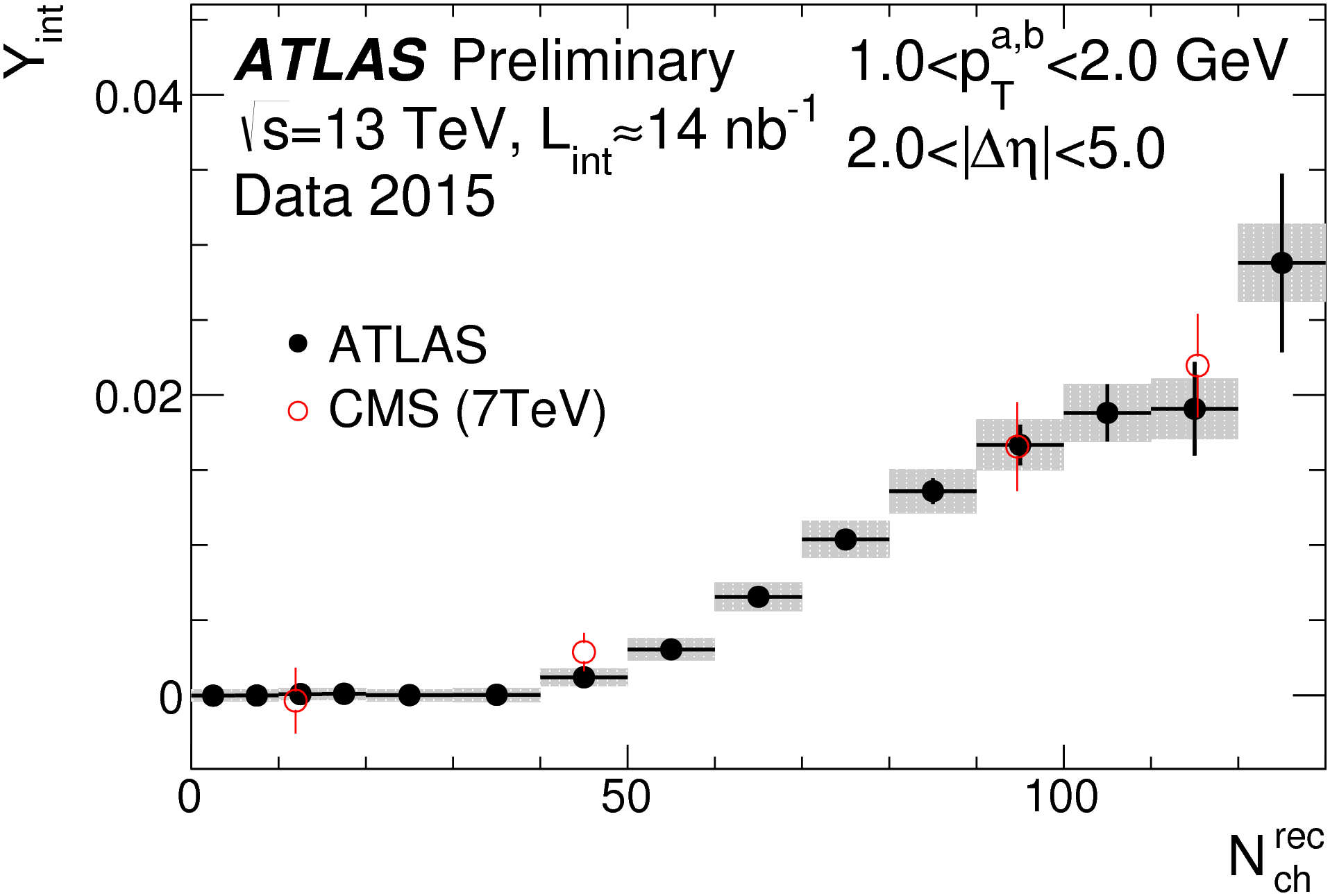}~~~
\includegraphics[width=0.47\textwidth]{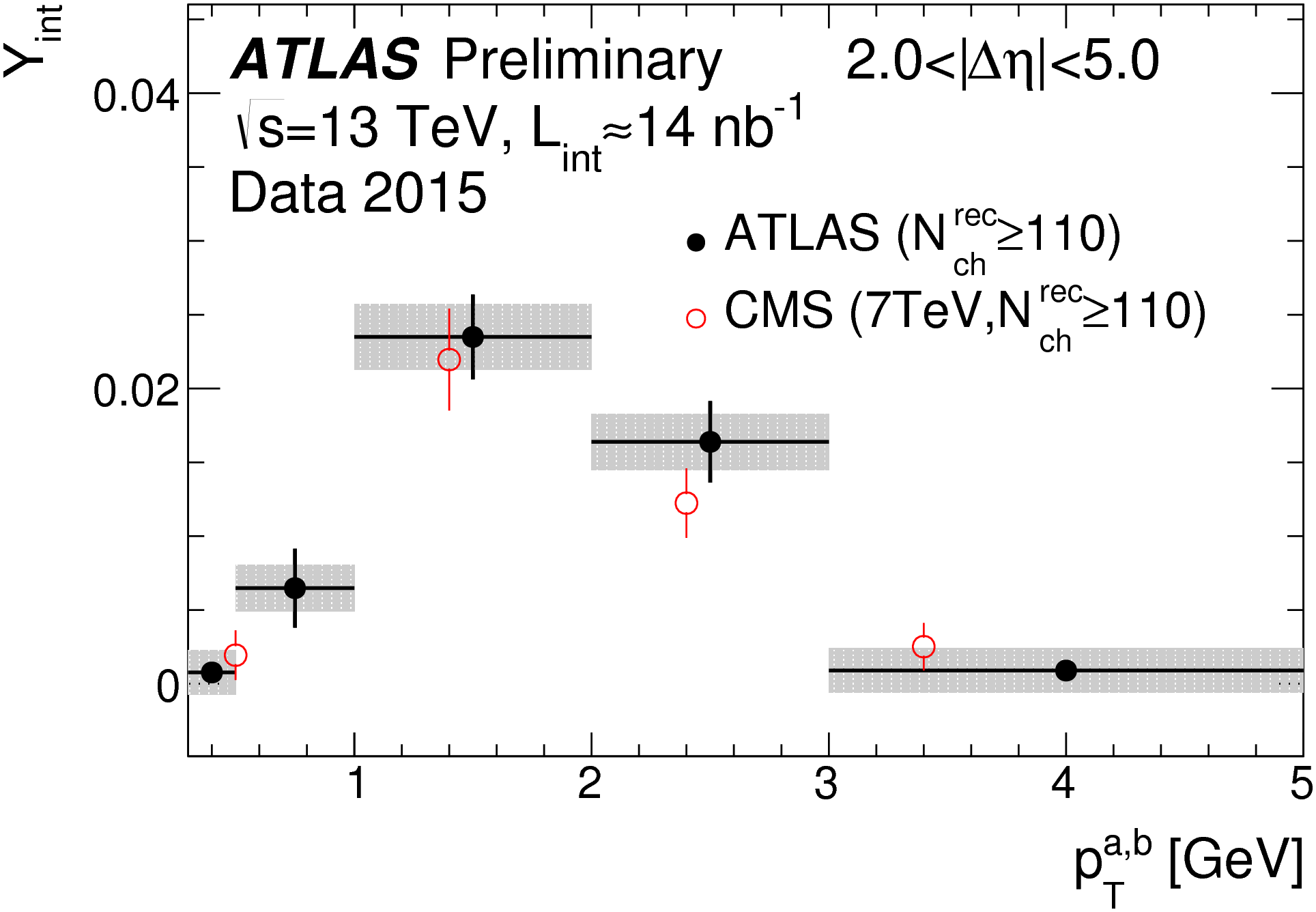}
\caption{The ridge yield in $\sqrt{s} = 13~\mbox{TeV}$ \pp{} collisions is shown as a function of multiplicity (left) and $\pta{}$ (right)~\cite{atlas13tev}. No significant dependence on the center-of-mass energy is observed.  (Note that these Preliminary results have been superseded by the results published in \cite{atlas13tevNew}.)\label{fig:atlas13tev}}
\end{figure}

\section{Conclusions}
The nearside and awayside ridge structures observed in two-particle correlations have been studied and characterized in measurements from ALICE, ATLAS, CMS, and LHCb.  The double ridge structure in \pPb{} collisions has been quantified by the parameter $v_2$, which shows a similar $\pta{}$-dependence as in \PbPb{}, as well as mass splitting like that observed in heavy-ion collisions.  Furthermore, it was observed that the $v_2$ is higher for particles in the Pb-going direction than in the p-going direction, when compared at the same relative multiplicity.  However, the ridge yield is identical in the positive- and negative-$\eta$ directions for the same absolute multiplicity.  These observations will serve to constrain future model calculations of  collective dynamics in p--A collisions.  The first measurements of the ridge in \pp{} collisions at $\sqrt{s} = 13~\mbox{TeV}$ show no significant difference from the results at $\sqrt{s} = 7~\mbox{TeV}$, and the first measurement of $v_2$ in \pp{} collisions has been performed.  The observation of long-range $\Delta\eta$ correlations in small collision systems at the LHC has significant implications for our understanding of the initial energy distribution of colliding protons (and larger nuclei) and the presence of collectivity over small length- and short time-scales.  

\nocite{*}
\bibliographystyle{aipnum-cp}%
\bibliography{biblio}%

\end{document}